\documentclass[prd,twocolumn,preprintnumbers,superscriptaddress,nofootinbib]{revtex4}
\pdfoutput=1

\usepackage{amsmath}
\usepackage{amssymb}
\usepackage{graphicx}
\usepackage{xspace}
\usepackage{epsfig}
\usepackage{units}
\usepackage{slashed}
\usepackage[hyperfootnotes=false]{hyperref}
\usepackage{gensymb}
\usepackage{cancel}
\usepackage{multirow}
\usepackage{float}
\usepackage[normalem]{ulem}

\newcommand{\eezh}{e^+ e^- \to ZH}
\allowdisplaybreaks

\begin{document}

\title{Fermionic Electroweak NNLO Corrections to \boldmath $e^+ e^- \to ZH$ with Polarized Beams and Different Renormalization Schemes}

\author{Ayres Freitas}
 \email{afreitas@pitt.edu}
\author{Qian Song}
 \email{qis26@pitt.edu}
\author{Keping Xie}
 \email{xiekeping@pitt.edu}
\affiliation{%
Pittsburgh Particle-physics Astro-physics \& Cosmology Center(PITT-PACC)\\
Department of Physics \& Astronomy, University of Pittsburgh, Pittsburgh, PA 15260, USA
}%

\begin{abstract}
Recently, the next-to-next-to-leading order (NNLO) electroweak corrections with fermion loops to the Higgsstrahling process were computed. Here we present numerical results for polarized electron/positron beams, as well as for two input parameter schemes known as the $\alpha(0)$ and $G_\mu$ schemes. The size of the NNLO corrections strongly depends on the beam polarization, leading to an increase of the $ZH$ cross-section by 0.76\% for $e^+_{\rm L} e^-_{\rm R}$ beams, and a decrease of 0.04\% for $e^+_{\rm R} e^-_{\rm L}$ beams. Furthermore, inclusion of the NNLO corrections is found to significantly reduce the discrepancy between the results in the $\alpha(0)$ and $G_\mu$ schemes. Using the remaining difference, together with other methods, the theory uncertainty from missing bosonic electroweak corrections is estimated to be less than 0.3\%. 
\end{abstract} 

\maketitle

\section{Introduction}

A high-luminosity $e^+e^-$ collider operating at center-of-mass energies of 240--250~GeV can perform model-independent precision measurements of the properties of the recently discovered Higgs boson. Several collider concepts have been proposed for this purpose: the International Linear Collider (ILC) \cite{ilc1,ilc2}, the Future Circular Collider (FCC-ee) \cite{fccee}, and the Circular Electron-Positron Collider (CEPC) \cite{cepc}. The cross-section for the main production process, $\eezh$, is projected to be measured with per-cent level precision at these facilities (1.2$\%$ at ILC, 0.4$\%$ at FCC-ee, and 0.5$\%$ at CEPC). Any deviations from the Standard Model (SM) expectation can be interpreted as sign of new physics beyond the SM.

Such an interpretation requires sufficiently accurate theoretical predictions for $ZH$ production within the SM, including higher-order radiative corrections. While the next-to-leading order (NLO) corrections have been known since many years \cite{nlo1,nlo2,nlo3}, the mixed electroweak-QCD NNLO corrections have been computed more recently \cite{ewqcd1,ewqcd2}. Very recently, the fermionic electroweak NNLO corrections have been completed \cite{Freitas:2022hyp}, where ``fermionic'' denotes contributions from diagrams with closed fermion loops. There is also ongoing work to compute the full electroweak NNLO corrections, including ``bosonic'' contributions \cite{Chen:2022mre}.

This article expands on the work of Ref.~\cite{Freitas:2022hyp} by including the effect of beam polarization and different renormalization schemes. Beam polarization can help to disentangle the nature of potential new physics effects in $ZH$ production. The SM NLO corrections to $\eezh$ with polarized beams have been presented in Ref.~\cite{Bondarenko:2018sgg}. Here, this is extended by also computing the fermionic electroweak NNLO corrections.

In addition, we present NNLO results for two different renormalization schemes. One scheme, called the $\alpha(0)$ scheme, uses the electromagnetic coupling, together with particle masses, as inputs to specify the SM parameters. This scheme has been used in Ref.~\cite{Freitas:2022hyp}. Alternatively, the $G_\mu$ scheme instead uses the Fermi constant as an input to define the electromagnetic coupling strength.
The numerical difference between the predictions in the two schemes can be taken as a proxy for the impact of missing higher-order corrections.

\section{Polarized cross-sections}

The perturbative expansion of the squared matrix element ${\cal M}$ takes the following form:
\begin{align}
    |{\cal M}|^2 =\, &|{\cal M}_{(0)}|^2 && \text{(LO)} \notag \\
    &+ 2\,\text{Re}\{{\cal M}_{(0)}^* {\cal M}_{(1)}\} && \text{(NLO)} \notag \\
    &+ |{\cal M}_{(1)}|^2 + 2\,\text{Re}\{{\cal M}_{(0)}^* {\cal M}_{(2)}\} && \text{(NNLO)},
\end{align}
where the subscript indicates the loop order. For the process $\eezh$, the matrix element has the form 
\begin{align}
{\cal M}_{(n)} = \bar{v}(p_{e^+})\Gamma^\mu_{(n)} u(p_{e^-})\,\varepsilon_{\mu} \,,
\end{align}
where $p_{e^\pm}$ are the momenta of the incoming electron and positron, respectively, and $\varepsilon_{\mu}$ is the polarization vector of the outgoing Z boson. For the unpolarized cross-section, one needs to average of initial spins, which is accomplished by the Carsimir trace technique:
\begin{align}
    \frac{1}{4}\sum_{e^\pm \text{ spins}} {\cal M}^*_{(m)} {\cal M}_{(n)} = \frac{1}{4}\,\text{Tr}\bigl\{\overline{\Gamma}^{\mu}_{(m)} \,\cancel{p}_{e^+}\,\Gamma_{\mu,(n)}\,\cancel{p}_{e^-}\bigr\},
\end{align}
where the electron mass has been neglected.

On the other hand, the results for left-/right-handed polarized electron and positron beams are obtained by inserting polarization projectors $P_{\rm R,L} = (1\pm\gamma^5)/2$:
\begin{align}
\begin{aligned}[b]
    {\cal M}^*_{(m)} {\cal M}_{(n)}\big|_{e^+_j e^-_k} = \frac{1}{4}\,\text{Tr}\bigl\{\overline{\Gamma}^{\mu}_{(m)}\,\cancel{p}_{e^+}\, P_j\, \Gamma_{\mu,(n)}\, P_k\, \cancel{p}_{e^-}\bigr\} & \\ (j,k=\text{L,R}), & 
\end{aligned}
\label{eq:polarizedM}\end{align}

After this step, the remaining calculation proceeds just in the same way as for the unpolarized cross-section. In particular, the non-trivial two-loop integrals can be computed with the techniques described in Refs.~\cite{Song:2021vru,Freitas:2022hyp}: One of the of the two subloops simplified using Feynman parameters and then expressed in terms of an dispersion integral. This allows one to solve the second subloop integral analytically in terms of well-known one-loop Passarino-Veltman functions. Ultraviolet (UV) divergencies are removed with suitable subtraction terms, which can be integrated analytically and then added back. Additional technical aspects of the computation of the subtraction terms are given in the appendix. Finally, the finite remainder integral is integrated numerically over the Feynman parameters and the dispersion variable. See Refs.~\cite{Song:2021vru,Freitas:2022hyp} for more details.

\medskip

Instead of separately going through all the steps of the computation for left- and right-handed polarized beams, one can alternatively derive the polarized matrix elements from the unpolarized one, which seems more efficient for calculating the two-loop diagrams. This is achieved by grouping certain types of diagrams. Considering a two-loop diagram where the incoming fermion line connects with $N$ gauge bosons($\gamma$ or $Z$), the polarized matrix elements satisfy the relationships
\begin{widetext}
\begin{align}
M_{(2)}^{V_1\cdots V_N*}M_{(0)}\big|_{e^+_Re^-_L}&=\frac{4\,g_{L}^{eeZ}\,g_{L}^{eeV_1}\cdots g_{L}^{eeV_N}}{g_{L}^{eeZ}\,g_{L}^{eeV_1}\cdots g_{L}^{eeV_N} + g_{R}^{eeZ}\,g_{R}^{eeV_1}\cdots g_{R}^{eeV_N}} \bigl[M_{(2)}^{V_1\cdots V_N*}M_{(0)}\bigr]_{\rm unpol}\,,~\notag \\
M_{(2)}^{V_1\cdots V_N*}M_{(0)}\big|_{e^+_Le^-_R}&=\frac{4\,g_{R}^{eeZ}\,g_{R}^{eeV_1}\cdots g_{R}^{eeV_N}}{g_{L}^{eeZ}\,g_{L}^{eeV_1}\cdots g_{L}^{eeV_N} + g_{R}^{eeZ}\,g_{R}^{eeV_1}\cdots g_{R}^{eeV_N}} \bigl[M_{(2)}^{V_1\cdots V_N*}M_{(0)}\bigr]_{\rm unpol}\,,~\notag \\[1ex]
M_{(2)}^{V_1\cdots V_N*}M_{(0)}\big|_{e^+_Le^-_L} &= M_{(2)}^{V_1\cdots V_N*}M_{(0)}\big|_{e^+_Re^-_R} = 0.
\end{align}
\end{widetext}
where $g_{L(R)}^{eeV}$ is the left(right)-handed coupling of $eeV$ vertex. 

If the fermion line connects only with W bosons, which only interact with left-handed fermions, thus we end up with  very simple equations
\begin{align}
M_{(2)}^{W_1\cdots W_N*}M_{(0)}\big|_{e^+_Re^-_L}&=4 \bigl[M_{(2)}^{W_1\cdots W_N*}M_{(0)}\bigr]_{\rm unpol}\,,~\notag \\
M_{(2)}^{W_1\cdots W_N*}M_{(0)}\big|_{e^+_Le^-_R}&=0.\end{align}

\medskip

\noindent
Numerical results for the polarized cross-section are presented in Tab.~\ref{tab:resPol}, using the following input parameters:
\begin{align}
& m_W^{\rm exp} = 80.379~\text{GeV} &&\Rightarrow\quad m_W = 80.352~\text{GeV}, \nonumber \\
&m_Z^{\rm exp} = 91.1876~\text{GeV}  &&\Rightarrow\quad m_Z = 91.1535~\text{GeV}, \nonumber \\[1ex]
&m_H = 125.1~\text{GeV}, &&  m_t = 172.76~\text{GeV}, \nonumber \\ 
&\alpha^{-1} = 137.036, && \Delta \alpha = 0.059, \nonumber \\
& \sqrt{s} = 240~\text{GeV}. \label{eq:input} \end{align}
where $\sqrt{s}$ represents the center-of-mass energy, and the masses of all the other fermions are set to be 0. 
Furthermore, $\Delta\alpha = 1-\alpha(m_Z)/\alpha(0)$ accounts for the running of the electromagnetic coupling between the Thomson limit and the weak scale, and it includes non-perturbative hadronic effects.

Since the Z and W bosons have sizeable decay widths, their masses ($m_Z$ and $m_W$) are defined via the complex pole of the gauge-boson propagators. This definition is theoretically well-defined but it differs from the one that is commonly used in experimental studies. The relation between the two is given by \cite{Bardin:1988xt}
\begin{align}
m_Z = m_Z^{\rm exp}\,[1+(\Gamma_Z^{\rm exp}/m_Z^{\rm exp})^2]^{-1/2}, \\ \qquad \Gamma_Z = \Gamma_Z^{\rm exp}\,[1+(\Gamma_Z^{\rm exp}/m_Z^{\rm exp})^2]^{-1/2}.
\label{massdef}
\end{align}
The numerical results presented in this paper do not include QED initial-state radiation (ISR). At the order that we are working, QED ISR factorizes and can be taken into account through convolution with a universal structure function.

\begin{table}
\begin{ruledtabular}
\begin{tabular}{lrr}
                &  $e^+_{\rm R} e^-_{\rm L}$ & $e^+_{\rm L} e^-_{\rm R}$ \\[.5ex]
\hline
$\sigma^\text{LO\phantom{NN}}$ [fb] & 541.28 & 350.55 \\
\hline
$\sigma^\text{NLO\phantom{N}}$ [fb] & 507.92 & 411.66 \\ 
\hline
$\sigma^\text{NNLO}$ [fb] & 507.51 & 418.68 \\[-.5ex]
$\qquad\mathcal{O}(\alpha^2_{N_f=2})$ & 1.75 & 5.77 \\[-.5ex]
$\qquad\mathcal{O}(\alpha^2_{N_f=1})$ & $-2.15$ & 1.25 \\ 
\end{tabular}
\end{ruledtabular}
\caption{Numerical results for the integrated $ZH$ production cross section, in fb, at LO, NLO and fermionic electroweak NNLO, for different beam polarizations. The results are for the $\alpha(0)$ renormalization scheme and $\sqrt{s}=240$~GeV. The electroweak NNLO corrections are also listed individually according to the number of fermion loops symbolized as $N_f$.}
\label{tab:resPol}
\end{table}

As can be seen from Tab.~\ref{tab:resPol}, the electroweak NNLO corrections depend strongly on the beam polarization. The contributions with two closed fermions loops ($N_f=2$) are significantly larger for right-handed electron polarization and left-handed positron polarization than for the opposite case. The contribution with one closed fermion loop ($N_f=1$) has opposite signs for the two polarization, which leads to an accidental cancellation for the unpolarized cross-section.

\begin{figure}
\includegraphics[width=\columnwidth]{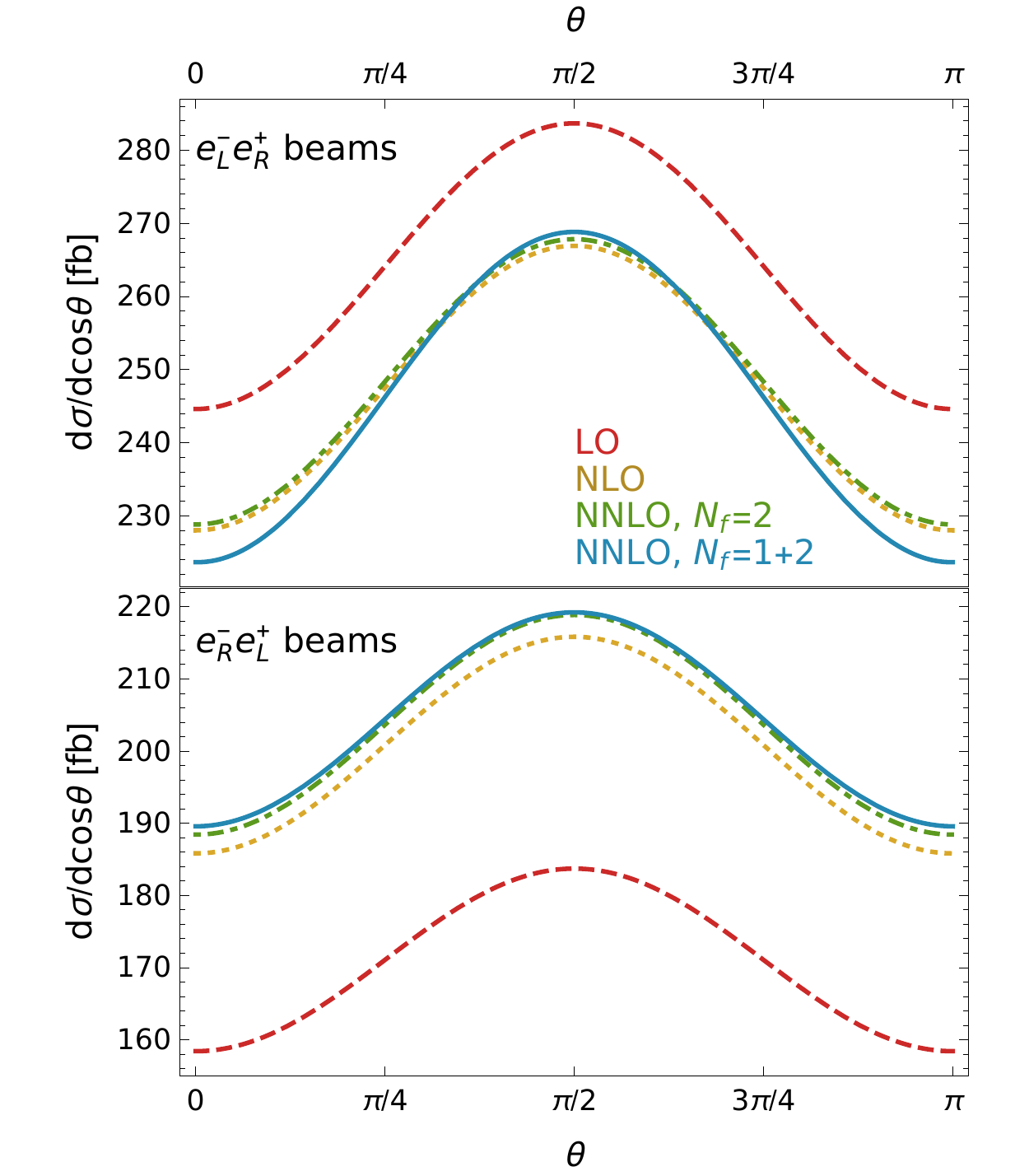}%
\vspace{-1ex}
\caption{Differential cross-sections, as a function of the scattering angle $\theta$, for two different beam polarizations. As in Tab.~\ref{tab:resPol}, the plots are based on the $\alpha(0)$ renormalization scheme and $\sqrt{s}=240$~GeV.}
\label{fig:resPol}
\end{figure}

The higher-order corrections also modify the shape of the differential cross-section, and the specific form of this modification depends on the beam polarization. This is illustrated in Fig~\ref{fig:resPol}, which shows the cross-section as a function of the scattering angle $\theta$ at different orders of perturbation theory for two beam polarizations. As can be seen from the figure, the shape distortions from the NNLO corrections are not very large, but non-negligible. They are more significant for the $e^-_{\rm L}e^+_{\rm R}$ polarization, which is plausible due to the impact of box diagrams with W bosons in this case.

\section{Results for different renormalization schemes}

The on-shell (OS) renormalization scheme defines the masses of all elementary particles via the (complex) propagator pole. In our calculation, this prescription is applied to the gauge boson masses, $m_{Z,W}$, the Higgs boson mass, $m_H$, and the top-quark mass, $m_t$. All other fermion masses are neglected. As already mentioned in the previous section, the finite width of the Z and W bosons needs to be taken into account, so that the poles of their propagators become complex. In deriving the $Z/W$ mass counterterms, the power counting $\Gamma_{Z,W}/m_{Z,W} \sim \mathcal{O}(\alpha)$ is used, \emph{i.e.} one performs a simultaneous expansion in the coupling constant and the gauge-boson widths. Since the Higgs-boson width is very small and the top quark only appear inside of loop contributions, their widths can be neglected. More details on the mass renormalization can be found, \emph{e.g.}, in Ref.~\cite{Freitas:2002ja}.

In addition, one needs a renormalization prescription for the electroweak coupling strength. In the following, two different schemes for this purpose are compared. One scheme, called the $\alpha(0)$ scheme, defines $\alpha=e^2/(4\pi)$ as the electromagnetic coupling at zero momentum, and the weak coupling is defined via
\begin{align}
    g = \frac{e}{\sin\theta_W} = \frac{e}{\sqrt{1-m_W^2/m_Z^2}} \label{eq:gcoupl}
\end{align}
to all orders in perturbation theory. This scheme is sensitive to the running of $\alpha(Q)$ from $Q=0$ to the weak scale, and it needs $\Delta\alpha$ as a numerical input.

The second scheme, called the $G_\mu$ scheme, relates the weak coupling to the Fermi coupling $G_\mu$,
\begin{align}
    \frac{G_\mu}{\sqrt{2}} = \frac{g^2}{8m_W^2}(1+\Delta r).
\end{align}
The numerical value for $G_\mu$ is extracted from the measured muon lifetime \cite{pdg}. The quantity $\Delta r$ contains radiative corrections that are determined by matching the muon decay matrix element in the Fermi theory and the full SM.  We use NLO and fermionic NNLO results for $\Delta r$ from Refs.~\cite{Freitas:2000gg,Freitas:2002ja} (also see Refs.~\cite{Awramik:2003ee}). The electromagnetic coupling in the $G_\mu$ scheme is derived from $g$ by again using eq.~\eqref{eq:gcoupl}. Note that this scheme does not depend on the shift $\Delta\alpha$ of the running electromagnetic coupling.

Using the input parameters in eq.~\eqref{eq:input}, together with $G_\mu = 1.1663787 \times 10^{-5}$, the results obtained in the two renormalization schemes are shown in Tab~\ref{tab:resScheme}. As can be seen from the table, the numerical agreement between the results in the two schemes improves with each order in perturbation theory, as expected.

\begin{table}
\begin{ruledtabular}
\begin{tabular}{lrr}
                &  $\alpha(0)$ scheme & $G_\mu$ scheme \\[.5ex]
\hline
$\sigma^\text{LO\phantom{NN}}$ [fb] & 222.96 & 239.18 \\
\hline
$\sigma^\text{NLO\phantom{N}}$ [fb] & 229.89 & 232.08 \\ 
\hline
$\sigma^\text{NNLO}$ [fb] & 231.55 & 232.74 \\[-.5ex]
$\qquad\mathcal{O}(\alpha^2_{N_f=2})$ & 1.88 & 0.73 \\[-.5ex]
$\qquad\mathcal{O}(\alpha^2_{N_f=1})$ & $-0.23$ & $-0.07$ \\ 
\end{tabular}
\end{ruledtabular}
\caption{Numerical results for the unpolarized integrated $ZH$ production cross section, in fb, for two different renormalization schemes. Results are given for $\sqrt{s}=240$~GeV at LO, NLO and fermionic electroweak NNLO. For the latter, the contributions from two ($N_f=2$) and one ($N_f=1$) closed fermion loops are also shown individually.}
\label{tab:resScheme}
\end{table}

In fact, this convergence is further improved when including the mixed electroweak-QCD two-loop corrections \cite{ewqcd1,ewqcd2}. We use numerical results for this contribution from Ref.~\cite{ewqcd2}. In order to do so, we have to compute our electroweak corrections for the same input parameters used there. The results are shown in Tab.~\ref{tab:withQCD}.

\begin{table}
\begin{ruledtabular}
\begin{tabular}{lrr}
                &  $\alpha(0)$ scheme & $G_\mu$ scheme \\[.5ex]
\hline
$\sigma^\text{LO}$ [fb]  & 223.14 & 239.64 \\
\hline
$\sigma^\text{NLO}$ [fb] & 229.78 & 232.46 \\ 
\hline
$\sigma^\text{NNLO,EW$\times$QCD}$ [fb] & 232.21 & 233.29 \\ 
\hline
$\sigma^\text{NNLO,EW}$ [fb] & 233.86 & 233.98 \\ 
\end{tabular}
\end{ruledtabular}
\caption{Similar to Tab.~\ref{tab:resScheme}, but using input values and mixed EW-QCD corrections from Ref.~\cite{ewqcd2}.}
\label{tab:withQCD}
\end{table}

The prediction for the cross-section including all available results agrees very well between the two renormalization schemes, with a difference of 0.12 fb. This difference is due to missing higher-order corrections, where the dominant impact is expected from the bosonic electroweak NNLO corrections, \emph{i.e.} from two-loop contributions without closed fermion loops.

Therefore, one can use the difference between the two renormalization schemes as an order-of-magnitude estimate of the perturbative theory uncertainty. Since this estimate is only a lower bound on the size of missing higher-order contributions, we conservatively multiply it by a factor 2, to arrive at an error estimate of 0.24~fb.

An alternative estimate of the bosonic NNLO corrections could be obtained by considering a subset of the latter, namely those stemming from $|{\cal M}_{(1,\rm bos)}|^2$, where ${\cal M}_{(1,\rm bos)}$ is the matrix element of the bosonic NLO corrections. This leads to a contribution of 0.65~fb to the cross-section. The contribution from genuine bosonic two-loop diagrams, $2\,\text{Re}\{{\cal M}_{(0)}^* {\cal M}_{(2,\rm bos)}\}$, is expected to be smaller than this, since the Born matrix element ${\cal M}_{(0)}$ contains several suppression factors: 
(a) the e-e-Z couplings in the initial state are smaller than the e-$\nu$-W couplings, which appear in the 1-loop box diagrams, by a factor $2^{-3/2} \sim 0.35$; (b) the s-channel Z propagator produces a factor $m_Z^2/(s-m_Z^2) \sim 0.17$ for $\sqrt{s}=240$~GeV.

Thus it seems plausible that the missing bosonic electroweak NNLO corrections have an impact between 0.24 and 0.65~fb on the SM prediction for the $ZH$ production cross-section. These theory error estimates are lower than the anticipated experimental precision (0.4--1\%), but a direct calculation of these missing contributions is still desirable.

\section{Conclusions}
In this article, we present the calculation of the $e^+e^-\to ZH$ cross section with polarized beams, while also addressing the renormalization scheme dependence. The electroweak NNLO corrections exhibit a strong dependence on the beam polarizations. The corrections are found to be large for $e_L^+e_R^-$ beam polarization, while small for $e_R^+e_L^-$ case due to numerical cancellation. By computing the cross section in the $\alpha(0)$ and $G_\mu$ schemes, we have shown that the renormalization scheme dependence decreases by including the two-loop electroweak corrections, and reduces further by adding mixed EW-QCD corrections. Renormalization scheme dependence can be utilized to estimate missing higher order corrections. Combining this with partial results for the missing bosonic electroweak NNLO corrections, we estimate the latter to be about $0.1-0.3\%$, thus lower than the anticipated experimental precision ($0.4-1\%$).

\section*{Acknowledgments}

This work has been supported in part by the National Science Foundation under grant no.~PHY-2112829.

\appendix
\section{UV subtraction terms}
\begin{figure}[t]
    \centering
    \includegraphics[scale=0.5]{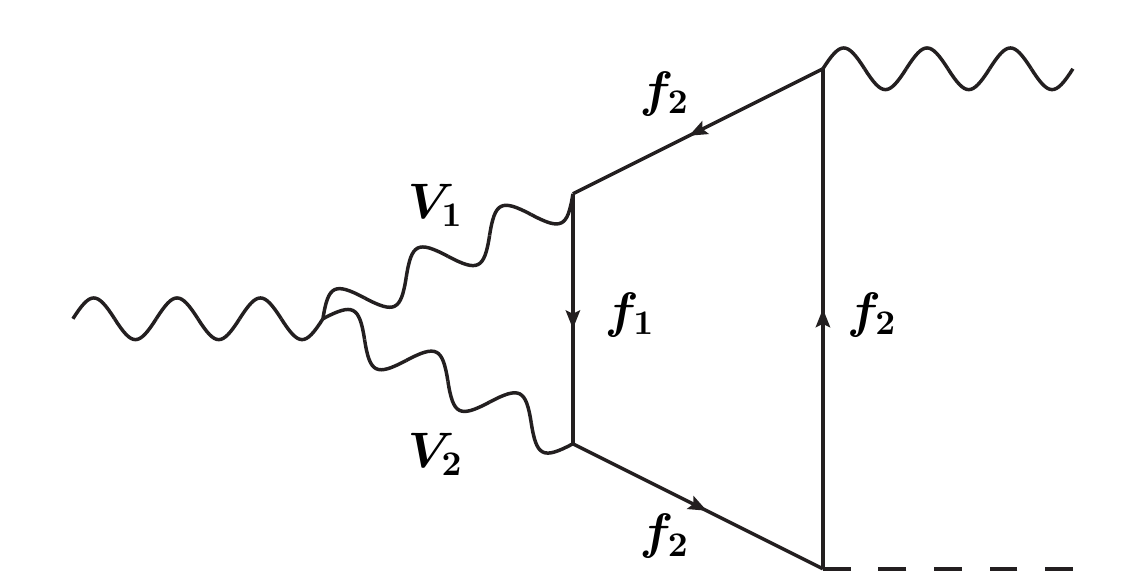}
    \caption{Two-loop VZH vertex diagram}
    \label{fig:2loopVZH}
\end{figure}
The UV divergent subtraction terms must be expanded appropriately to get the correct finite term. As stated in Ref.~\cite{Freitas:2022hyp}, there are three types of subtraction terms: two for subloop divergences and one more for a global divergence. The latter corresponds to vacuum diagrams, the analytical formulas of which can be obtained and expanded to higher orders in $\epsilon$ with TVID \cite{tvid}. The general form for the subloop divergences can be expressed as the multiplication of two one-loop scalar integrals. Taking the two-loop vertex diagram Fig.~\ref{fig:2loopVZH} as an example, the tensor integral can be written as
\begin{align}
I &=\int \frac{d^D q_2}{i\pi^2} \frac{d^D q_1}{i\pi^2} \sum_{n_0,n_1,n_2,i,j} c_{ij}^{n_0,n_1,n_2} \times \{p_i^{n_0},q_1^{n_1},q_2^{n_2}\}_j \notag \\
&\times \frac{1}{(q_2^2-m_{V_2}^2)((q_2+p)^2-m_{V_1}^2)((q_2+q_1)^2-m_{f_1}^2)} \notag \\
&\times\frac{1}{(q_1^2-m_{f_2}^2)((q_1-p_h)^2-m_{f_2}^2)((q_1-p)^2-m_{f_2}^2)} \,,
\label{eq:int1}
\end{align}
where $\{p_i^{n_0},q_1^{n_1},q_2^{n_2}\}$ denotes dot products among external momentum $p_i$ and loop momentum $q_{1,2}$, and $n_i$ denote the power of each of them. The index $j$ labels all possible dot product conditions. For example for $n_0=0,n_1=n_2=2$, the possible dot products read
\begin{align}
&\{p_i^{0},q_1^{2},q_2^{2}\}_1=(q_1\cdot q_1)(q_2\cdot q_2),\notag\\
&\{p_i^{0},q_1^{2},q_2^{2}\}_2=(q_1\cdot q_2)(q_1\cdot q_2).
\end{align}
The SM Feynman rules require that $n_1\leq 4, n_2\leq 2,n_0+n_1+n_2\leq 6$. $c_{ij}^{n_0,n_1,n_2}$ is the coefficient of a dot product, and it is a function of masses and dimension $D$. The  integral \eqref{eq:int1} contains a subloop divergence from the $q_1$ loop, which originate from the numerators $q_1^{n_1}$ with $n_1\geq 4$. To make the $q_1$ integral UV finite, the following subtraction term is constructed:
\begin{align}
I_{\text{subtr}}^{q_1}& = \int \frac{d^D q_2}{i\pi^2} \,\frac{d^D q_1}{i\pi^2} \sum_{i,j} \Big[ c_{ij}^{2,4,0} \times \{p_i^2,q_1^4,q_2^0\}_j \notag\\
&+c_{ij}^{1,4,1} \times \{p_i^1,q_1^4,q_2^1\}_j +c_{ij}^{0,4,2} \times \{p_i^0,q_1^4,q_2^2\}_j \notag \\
&+c_{ij}^{1,4,0} \times \{p_i^1,q_1^4,q_2^0\}_j +c_{ij}^{0,4,1} \times \{p_i^0,q_1^4,q_2^1\}_j \notag\\
&+ c_{ij}^{0,4,0} \times \{p_i^0,q_1^4,q_2^0\}_j \Big]\notag \\
&\times \frac{1}{(q_2^2-m_{V_2}^2)((q_2+p)^2-m_{V_1}^2)(q_1^2-m_{f_1}^2)} \notag \\
&\times\frac{1}{(q_1^2-m_{f_2}^2)(q_1^2-m_{f_2}^2)(q_1^2-m_{f_2}^2)} \,.
\label{eq:q1subtra1} 
\end{align}
From Eq.~\ref{eq:q1subtra1}, one can see that the loop integrals of $q_1$ and $q_2$ are disentangled. After performing the loop integration, one obtains
\begin{align}
I_\text{subtr}^{q_1} =  B_0(p^2,m_{V_2}^2,m_{V_1}^2) \times \Big[ a_1 A_0(m_{f_1}^2) + a_2A_0(m_{f_2}^2) \Big] \label{eq:q1subtra2} 
\end{align}
where $a_i$ are functions of masses, external momenta and dimension $D$. A similar subloop subtraction term needs to be introduced in the vacuum integrals for the global divergence.
Combining the two subloop subtraction terms, we obtain
\begin{align}
I_\text{subtr} &=  \Big[ B_0(p^2,m_{V_2}^2,m_{V_1}^2)-B_0(0,m_{V_2}^2,m_{V_1}^2) \Big] \notag \\
&\quad\times \Big[ a_1 A_0(m_{f_1}^2) + a_2A_0(m_{f_2}^2) \Big]
\end{align}
This term can now be expanded in powers of \mbox{$\epsilon=(4-D)/2$}, resulting in the expressions
\begin{align}
I_\text{subtr}^{\text{div}} &=  \Big[ B_0^{(0)}(p^2,m_{V_2}^2,m_{V_1}^2)-B_0^{(0)}(0,m_{V_2}^2,m_{V_1}^2) \Big] \notag \\
&\quad\times \Big[ a_1^{(0)} A_0^{(-1)}(m_{f_1}^2) + a_2^{(0)} A_0^{(-1)}(m_{f_2}^2) \Big], \\[1ex]
I_\text{subtr}^{\text{fin}} &= \Big[ B_0^{(0)}(p^2,m_{V_2}^2,m_{V_1}^2)-B_0^{(0)}(0,m_{V_2}^2,m_{V_1}^2) \Big] \notag \\
&\quad\times \Big[ a_1^{(0)} A_0^{(0)}(m_{f_1}^2) + a_2^{(0)} A_0^{(0)}(m_{f_2}^2) \notag\\
&\qquad+a_1^{(1)} A_0^{(-1)}(m_{f_1}^2) + a_2^{(1)} A_0^{(-1)}(m_{f_2}^2) \Big] \notag\\
&+\Big[ B_0^{(1)}(p^2,m_{V_2}^2,m_{V_1}^2)-B_0^{(1)}(0,m_{V_2}^2,m_{V_1}^2) \Big] \notag \\
&\quad\times\Big[ a_1^{(0)} A_0^{(-1)}(m_{f_1}^2) + a_2^{(0)} A_0^{(-1)}(m_{f_2}^2) \Big]. \label{eq:q1subtra4}
\end{align}
where $(n)$ denote the expansion order in $\epsilon$. Eq.~\ref{eq:q1subtra4} indicates that $\mathcal{O}(\epsilon)$ parts of one-loop scalar functions must be taken into account. Analytical expressions for these can be found in Ref.~\cite{Nierste:1992wg}.


\bibliographystyle{utcaps_mod}
\bibliography{mollerbib}

\begin{thebibliography}{99}
\bibitem{ilc1}
H.~Baer \textit{et al.}
``The International Linear Collider Technical Design Report - Volume 2: Physics,''
[arXiv:1306.6352 [hep-ph]].

\bibitem{ilc2}
P.~Bambade \textit{et al.}
``The International Linear Collider: A Global Project,''
[arXiv:1903.01629 [hep-ex]].

\bibitem{fccee}
A.~Abada {\it et al.} [FCC Collaboration],
``FCC-ee: The Lepton Collider : Future Circular Collider Conceptual Design Report Volume 2,''
Eur.\ Phys.\ J.\ ST {\bf 228}, 261 (2019).
  
\bibitem{cepc}
J.~B.~Guimar\~aes da Costa {\it et al.} [CEPC Study Group],
``CEPC Conceptual Design Report: Volume 2 - Physics \& Detector,''
arXiv:1811.10545 [hep-ex].

\bibitem{nlo1}
J.~Fleischer and F.~Jegerlehner,
``Radiative Corrections to Higgs Production by $e^+ e^- \to Z H$ in the \{Weinberg-Salam\} Model,''
Nucl. Phys. B \textbf{216}, 469 (1983).
\bibitem{nlo2}
B.~A.~Kniehl,
``Radiative corrections for associated $Z H$ production at future $e^{+} e^{-}$ colliders,''
Z. Phys. C \textbf{55}, 605 (1992).
\bibitem{nlo3}
A.~Denner, J.~K\"ublbeck, R.~Mertig and M.~B\"ohm,
``Electroweak radiative corrections to $e^+ e^- \to Z H$,''
Z. Phys. C \textbf{56}, 261 (1992).


\bibitem{ewqcd1}
Y.~Gong, Z.~Li, X.~Xu, L.~L.~Yang and X.~Zhao,
``Mixed QCD-EW corrections for Higgs boson production at $e^+e^-$ colliders,''
Phys. Rev. D \textbf{95}, 093003 (2017)
[arXiv:1609.03955 [hep-ph]].
\bibitem{ewqcd2}
Q.~F.~Sun, F.~Feng, Y.~Jia and W.~L.~Sang,
``Mixed electroweak-QCD corrections to e+e-\textrightarrow{}HZ at Higgs factories,''
Phys. Rev. D \textbf{96}, no.5, 051301(R) (2017)
[arXiv:1609.03995 [hep-ph]].


\bibitem{Freitas:2022hyp}
A.~Freitas and Q.~Song,
``Two-Loop Electroweak Corrections with Fermion Loops to $e+e-\rightarrow{}ZH$,''
Phys. Rev. Lett. \textbf{130}, no.3, 031801 (2023)
[arXiv:2209.07612 [hep-ph]].

\bibitem{Chen:2022mre}
X.~Chen, X.~Guan, C.~Q.~He, Z.~Li, X.~Liu and Y.~Q.~Ma,
``Complete two-loop electroweak corrections to $e^+e^-\rightarrow HZ$,''
[arXiv:2209.14953 [hep-ph]].


\bibitem{Bondarenko:2018sgg}
S.~Bondarenko, Y.~Dydyshka, L.~Kalinovskaya, L.~Rumyantsev, R.~Sadykov and V.~Yermolchyk,
``One-loop electroweak radiative corrections to polarized $e^+e^- \to ZH$,''
Phys. Rev. D \textbf{100}, 073002 (2019).
[arXiv:1812.10965 [hep-ph]].

\bibitem{Song:2021vru}
Q.~Song and A.~Freitas,
``On the evaluation of two-loop electroweak box diagrams for $e^+e^- \to HZ$ production,''
JHEP \textbf{04}, 179 (2021)
[arXiv:2101.00308 [hep-ph]].

\bibitem{Bardin:1988xt}
  D.~Y.~Bardin, A.~Leike, T.~Riemann and M.~Sachwitz,
  ``Energy Dependent Width Effects in e+ e- Annihilation Near the Z Boson Pole,''
  Phys.\ Lett.\ B {\bf 206}, 539 (1988).

\bibitem{Freitas:2002ja}
A.~Freitas, W.~Hollik, W.~Walter and G.~Weiglein,
``Electroweak two loop corrections to the $M_W-M_Z$ mass correlation in the standard model,''
Nucl. Phys. B \textbf{632}, 189-218 (2002)
[erratum: Nucl. Phys. B \textbf{666}, 305-307 (2003)]
[arXiv:hep-ph/0202131 [hep-ph]].

\bibitem{pdg}
R.~L.~Workman \textit{et al.} [Particle Data Group],
``Review of Particle Physics,''
PTEP \textbf{2022}, 083C01 (2022).

\bibitem{Freitas:2000gg}
A.~Freitas, W.~Hollik, W.~Walter and G.~Weiglein,
``Complete fermionic two loop results for the $M_W$--$M_Z$ interdependence,''
Phys. Lett. B \textbf{495}, 338-346 (2000)
[erratum: Phys. Lett. B \textbf{570}, no.3-4, 265 (2003)]
[arXiv:hep-ph/0007091 [hep-ph]].

\bibitem{Awramik:2003ee}
M.~Awramik and M.~Czakon,
``Complete two loop electroweak contributions to the muon lifetime in the standard model,''
Phys. Lett. B \textbf{568}, 48-54 (2003)
[arXiv:hep-ph/0305248 [hep-ph]].

\bibitem{tvid}
S.~Bauberger, A.~Freitas and D.~Wiegand,
``TVID 2: Evaluation of planar-type three-loop self-energy integrals with arbitrary masses,''
JHEP \textbf{01}, 024 (2020)
[arXiv:1908.09887 [hep-ph]].

\bibitem{Nierste:1992wg}
U.~Nierste, D.~Muller and M.~Bohm,
``Two loop relevant parts of D-dimensional massive scalar one loop integrals,''
Z. Phys. C \textbf{57}, 605-614 (1993).

\end{thebibliography}

\end{document}